\documentclass[aps,pra,reprint,floatfix,amsmath,amssymb,superscriptaddress,
]{revtex4-1}
\usepackage[utf8]{inputenc}
\usepackage{textcomp}
\usepackage{amsmath}
\usepackage{amssymb}
\usepackage{graphicx}
\usepackage{braket}
\usepackage{url} 

\begin{abstract} \begin{center}\textbf{Abstract}\end{center} When a well-localized photon is incident on a spatially superposed absorber but is not absorbed, the photon can still deliver energy to the absorber.  It is shown that when the transferred energy is small relative to the energy uncertainty of the photon, this constitutes an unusual type of weak measurement of the absorber's energy, where the energy distribution of the unabsorbed photon acts as the measurement device, and the strongly disturbed state of the absorber becomes the effective pre-selection.  Treating the final state of the absorber as the post-selection, it is shown that the absorber's energy increase is the weak value of its translational Hamiltonian, and the energy distribution of the photon shifts by the opposite amount.  The basic case of non-scattering is examined, followed by the case of interaction-free energy transfer.  Details and interpretations of the results are discussed.
\end{abstract}

\begin{document}

\title{Energy-Based Weak Measurement}

\author{Mordecai Waegell}
\affiliation{Institute for Quantum Studies, Chapman University, Orange, CA 92866, USA}
\affiliation{Schmid College of Science and Technology, Chapman University, Orange, CA 92866, USA}

\author{Cyril Elouard}
\affiliation{Department of Physics and Astronomy, University of Rochester, Rochester, New York  14627, USA}

\author{Andrew N. Jordan}
\affiliation{Institute for Quantum Studies, Chapman University, Orange, CA 92866, USA}
\affiliation{Department of Physics and Astronomy, University of Rochester, Rochester, New York  14627, USA}

\date{\today}

\maketitle

\section{Introduction}
This conference proceedings article is a follow-up on the presented research \cite{elouard2019spooky}.

Weak measurement of quantum systems is a subtle area of metrology where a system is measured in such a way that very little information is gained by measurement device, while the state of the system is barely disturbed, and collapses only rarely.  A weak measurement, followed by a projective measurement of the system, induces interference in the measurement device, causing the average readout of the device to be given by the weak value of the weakly measured system observable.  The weak value of an observable, $\hat{A}$, first introduced by Aharonov, Albert, and Vaidman \cite{aharonov1988result}, is a complex quantity defined using both the prepared state (called the pre-selection, $\ket{\psi}$) and the state obtained from the final projective measurement (the post-selection, $\bra{\phi}$) as,

\begin{equation}
    \hat{A}_w \equiv \frac{\bra{\phi}\hat{A}\ket{\psi}}{\langle\phi|\psi\rangle}
\end{equation}

In most measurements, a known interaction Hamiltonian couples the system to the measurement device (henceforth called the pointer), causing their degrees of freedom to become entangled.  However, if the form of the interaction Hamiltonian is not clearly known, sometimes the measurement can still be clearly understood by considering quantities that are conserved during the interaction.

Here, we examine such a case, where the pointer is the broad energy distribution $\phi(\omega)$ of a single well-localized photon, which couples via energy conservation to the translational state of a perfect absorber, even when it is not absorbed\footnote{Formally, our analysis applies to any object which perfectly absorbs the photon, or scatters it out of its original path, but we have chosen to use `absorber' throughout the article for simplicity.}.  Although we do not know the form of the operator which induces translations in the energy distribution of the photon, we still find that upon post-selection, the pointer encodes a weak value, which in this case corresponds to the energy it exchanged with the absorber.  The subtlety of this scenario lies in understanding the nature of the coupling that occurs between the photon and the absorber, and identifying from it both the pre-selection and the observable that is being weakly measured.

We consider a simple scenario where the photon is simply incident on the absorber, and then we move to the more interesting case where the photon performs an interaction-free measurement of the absorber, and the energy appears to be transferred to it nonlocally.

\section{Energy Conservation During Coherent Non-Scattering}

It was observed by Dicke \cite{dicke1981interaction} that when a localized photon is incident on an absorber in a larger spatial superposition, energy is exchanged between the photon and the absorber, even when the photon is not absorbed.  This effectively happens because after the photon has passed, the absorber's wavefunction has been reshaped to have a much lower probability in the target region, which generally raises its average energy.  Technically speaking, the photon and absorber enter an entangled state through a process of repeated absorption and emission, and the absorber gains translational energy from the photon through this process.

We consider the simplified case of a perfect absorber, which always absorbs the photon if they are found in the same place, and whose translational state is modeled as a 2-level system.  The absorber begins in its translational ground state, $|g\rangle = \alpha|\textrm{in}\rangle + \beta|\textrm{out}\rangle$, where $|\textrm{in}\rangle$ and $|\textrm{out}\rangle$ are positions of the absorber inside and outside of the path of the photon, respectively (and of course $|\alpha|^2 + |\beta|^2 = 1$).  We set the ground state energy to zero, and the energy of the first excited state $|f\rangle = \beta^*|\textrm{in}\rangle - \alpha^*|\textrm{out}\rangle$ to $\omega_f$, so that the absorber's translational Hamiltonian is simply $\hat{H} = \omega_f |f\rangle\langle f|$.   

The entire envelope of the photon path is incident on the position $|\textrm{in}\rangle$.  The initial state of the photon and absorber is,
\begin{equation}
    \int d\omega\big(  \alpha|\textrm{in}\rangle + \beta|\textrm{out}\rangle \big) \phi(\omega) a^\dag_\omega |0\rangle ,\label{Initial}
\end{equation}
where $\phi(\omega)$ is the energy wavefunction of the photon, $|0\rangle$ is the vacuum state of the photon field, and $a^\dag_\omega$ is the creation operator for a photon of energy $\omega$. With probability $|\alpha|^2$, the absorber is in the photon's path and absorbs it.  With probability $|\beta|^2$, the absorber is outside the photon's path, and the photon is not absorbed.  Then, applying a special energy-conserving projection of Eq. \ref{Initial} onto the absorber state $|\textrm{out}\rangle$ results in the (unnormalized) state,
\begin{equation}
    \int d\omega  \big(  \beta^*\phi(\omega)|g\rangle - \alpha\phi(\omega + \omega_f)|f\rangle \big)a^\dag_\omega |0\rangle,\label{eq:2}
\end{equation}
where $|\textrm{out}\rangle$ has been expanded into the energy eigenbasis, and each term shows the exchange of a definite amount of energy between the photon and absorber such that their average energy is conserved term by term.  The technical details of this local interaction between the photon and the translational energy eigenstates of the absorber are quite subtle, and were developed in \cite{elouard2019spooky}.

The strength of the measurement is determined by the ratio $\omega_f/\sigma$, where $\sigma$ is the standard deviation of the pointer distribution $\phi(\omega)$.  When $\omega_f/\sigma \gg 1$, Eq. \ref{eq:2} is a strongly entangled state, which is the case of a strong projective measurement, where a single measurement of the pointer energy projects the absorber onto an energy eigenstate.  Conversely, when $\omega_f/\sigma \ll 1$, Eq. \ref{eq:2} is a weakly entangled state, which is the case of a weak measurement, where a single measurement of the pointer energy barely disturbs the original state of the absorber.

Even though it all happens at once, it is instructive to think of the measurement interaction as occurring in two logical steps.  First, the absorber state is strongly  projected to $|\textrm{out}\rangle$ by the presence of the unabsorbed photon, which becomes its pre-selection $|\psi\rangle$, and the photon's energy distribution $\phi(\omega)$ remains unchanged.  Second, each energy eigenstate in $|\textrm{out}\rangle$ couples to $\phi(\omega)$ by producing a shift equal to its energy eigenvalue, resulting in the entangled state of Eq. \ref{eq:2}.  We might imagine this energy shift as generated by a virtual `time' operator, complementary to the photon's Hamiltonian.

Working in the weak regime, if we measure the (in,out) basis we find $|\textrm{out}\rangle$ with certainty (to first order in $\omega_f/\sigma$), and we take this to be our post-selection for the absorber for this analysis (specifically because measuring this basis does not change the state). The weak value of the negative Hamiltonian of the absorber is then,
\begin{equation}
 -\hat{H}_w = -\langle\hat{H}\rangle =  -\omega_f|\langle \textrm{out} | f \rangle|^2 = -\omega_f|\alpha|^2,
\end{equation}
 where the negative sign ensures conservation of energy between the two systems.
Projecting Eq. \ref{eq:2} onto $|\textrm{out}\rangle$ results in the interfering energy distribution,

\begin{equation}
|\alpha|^2 \phi(\omega + \omega_f) + |\beta|^2 \phi(\omega) \approx \phi(\omega + \omega_f|\alpha|^2), \label{Measured}
\end{equation}
where the approximation is also first order in $\omega_f/\sigma$.  

Thus, we see that in the weak regime, the shift in the energy distribution of the unabsorbed photon is the negative weak value of the absorber's Hamiltonian, and thus it has served as the pointer for a weak measurement --- in this case, a measurement of how much energy was transferred from the photon to the absorber.

A very important subtlety of this analysis is that the absorber must actually begin in the ground state, even though we take the pre-selection to be the state of the absorber conditioned on the photon not being absorbed.  If the initial state of the absorber was actually $|\textrm{out}\rangle$, then the photon would simply miss, and it would exchange no energy with the absorber at all.  The energy exchange occurs because the photon must do local work on the absorber in order to suppress the $\ket{\textrm{in}}$ component of the ground state.  

\section{Interaction-Free Energy Transfer}

Now we move on to a more interesting case, which we recently introduced \cite{elouard2019spooky}, where the photon appears to transfer energy nonlocally to the absorber as it passes through a Mach-Zehnder interferometer (MZ).  The MZ is tuned so that a photon entering from one port always exits at a single port (called the bright port) due to constructive interference, and never at the other exit port (called the dark port) due to destructive interference.  If we introduce an obstruction on one arm of the MZ, which blocks the photon, then the interference is lost, and it becomes possible for a photon traveling the other arm to exit the dark port.  When this happens, the photon has detected the presence of the obstruction on one arm, without ever visiting that arm, giving rise to the name \textit{interaction-free measurement} \cite{elitzur1993quantum}.

Let us now consider the case where instead of a concrete (classical) obstruction, we place the absorber of the previous section into arm I of the MZ, where its ground state is again a superposition of being inside and outside the photon's path on that arm.

Once the photon has passed through the first beam splitter of the MZ, the state of the photon and absorber is,
\begin{equation}
    \frac{1}{\sqrt{2}}\int d\omega  \big(  \alpha|\textrm{in}\rangle + \beta|\textrm{out}\rangle \big)\phi(\omega)\big(a^\dag_{\omega,\textrm{I}} + a^\dag_{\omega,\textrm{II}}\big) |0\rangle,
\end{equation}
where the creation operators are now indexed to create a photon of frequency $\omega$ on a specific arm (I or II) of the MZ.  With probability $|\alpha|^2/2$ the photon is on arm I and the absorber is in its path and absorbs it.  With probability $|\beta|^2/2$ the photon is on arm I and the absorber is outside its path and does not absorb it.  And with probability $1/2$ the photon is on arm II of the MZ and never goes near the absorber.

If the photon is not absorbed, the (unnormalized) state is then,
\begin{equation}
    \int d\omega  \Big[ \big(|\beta|^2\phi(\omega)|g\rangle - \alpha \beta\phi(\omega+\omega_f)|f\rangle    \big)a^\dag_{\omega,\textrm{I}} +  \phi(\omega)|g\rangle a^\dag_{\omega,\textrm{II}} \Big] |0\rangle, \label{Mid}
\end{equation}
where we have switched back to the energy eigenbasis of the absorber.  Next the photon passes through the second beam splitter of the MZ, which is characterized by the transformation,
\begin{equation}
a^\dag_{\omega,\textrm{I}} \rightarrow \frac{1}{\sqrt{2}}\big( a^\dag_{\omega,\textrm{Br}}   + a^\dag_{\omega,\textrm{Dk}}  \big),
\end{equation}
\begin{equation}
a^\dag_{\omega,\textrm{II}} \rightarrow \frac{1}{\sqrt{2}}\big( a^\dag_{\omega,\textrm{Br}}   - a^\dag_{\omega,\textrm{Dk}}  \big), \nonumber
\end{equation}
onto the bright (Br) and dark (Dk) exit ports of the MZ. Collecting the terms for each exit port we have,

\begin{equation}
    \int d\omega  \Big( \big[(|\beta|^2+1)\phi(\omega)|g\rangle - \alpha \beta\phi(\omega+\omega_f)|f\rangle    \big]a^\dag_{\omega,\textrm{Br}} \label{Final}
\end{equation}
\begin{equation}
    + \big[(|\beta|^2-1)\phi(\omega)|g\rangle - \alpha \beta\phi(\omega+\omega_f)|f\rangle    \big]a^\dag_{\omega,\textrm{Dk}} \Big) |0\rangle.  \nonumber
\end{equation}
As in the case of the classical obstruction, the presence of the quantum absorber makes it possible for the photon to exit the dark port of the MZ, and as we will show, this constitutes an interaction free measurement in much the same way.

We still want to work in the weak regime where $\omega_f \ll \sigma$, which also means that the energy shift does not significantly decrease the visibility of the interference at the second beam splitter.  Taking the approximation that $\phi(\omega) \approx \phi(\omega+\omega_f)$, we see that when the photon exits the dark port, the absorber is left in the state $|\textrm{in}\rangle = \alpha^*|g\rangle + \beta|f\rangle$ --- the state where it would have definitely absorbed a photon on arm I.  As a result, the average energy of the absorber has increased by $\omega_f|\beta|^2$, but had the photon taken  arm I it would have been absorbed and never reached the dark port, so this appears to be an \textit{interaction-free energy transfer}.

It is important to note that this energy transfer only appears nonlocal if one retrodicts  a single classical trajectory that the photon must have followed through the MZ, for the outcome that obtained.

However, in the formal treatment, the interaction Hamiltonian transfers energy locally to the photon in arm I, and it is the wave interference at the second beam splitter, of the photon terms coming from both arms, which completes the energy transfer to the photon which finally escapes the dark port.

Now, to return to the issue of weak measurement.  As we will show, the photon performs a weak measurement of the absorber's energy on arm I of the MZ.  The measured observable is $\hat{A} = -|\textrm{I}\rangle \langle \textrm{I}| \hat{H}$, where $\hat{H} = \omega_f|f\rangle\langle f|$ is the translational Hamiltonian of the absorber, and $|\textrm{I}\rangle \langle \textrm{I}|$ is the projector onto the photon being on arm I, using the compact definitions, $|\textrm{I}\rangle = \int d\omega \phi(\omega) a^\dag_{\omega,\textrm{I}}|0\rangle$ and $|\textrm{II}\rangle = \int d\omega \phi(\omega) a^\dag_{\omega,\textrm{II}}|0\rangle$.  This gives us,
\begin{equation}
\hat{A} = - \omega_f |\textrm{I}\rangle\langle\textrm{I}|\otimes|f\rangle\langle f|.  
\end{equation}

Now, if we break the interaction into two steps as in the previous section, then first the presence of the unabsorbed photon strongly projects the state onto the pre-selection,
\begin{equation}
|\psi\rangle =  \beta|\textrm{out}\rangle |\textrm{I}\rangle  + \alpha|\textrm{in}\rangle|\textrm{II}\rangle + \beta|\textrm{out}\rangle|\textrm{II}\rangle,
\end{equation}
and second, the energy eigenstates in $|\psi\rangle$ couple weakly to $\phi(\omega)$ by producing shifts equal to their energy eigenvalues, resulting in the entangled state of Eq. \ref{Mid}.

If we measure the absorber in the (in,out) basis after the photon has exited the dark port, we find the state is $|\textrm{in}\rangle$ with certainty (to first order), and thus we will take our postselection at the exit port to be $|\textrm{in}\rangle|\textrm{Dk}\rangle$. Retropropagating this back through the second beam splitter gives us,
\begin{equation}
|\phi\rangle =|\textrm{in}\rangle \big(|\textrm{I}\rangle - |\textrm{II}\rangle\big),
\end{equation}
from which we obtain the weak value,
\begin{equation}
\hat{A}_w = -\omega_f|\beta|^2.
\end{equation}

Finally, projecting Eq. \ref{Final} directly onto the case where the photon exits the dark port and the absorber is left in the state $|\textrm{in}\rangle$, we obtain the interfering energy distribution,
\begin{equation}
\phi(\omega) - |\beta|^2 \phi(\omega) + |\beta|^2 \phi(\omega + \omega_f) \approx \phi(\omega + \omega_f|\beta|^2),
\end{equation}
where the approximation is first order in $\omega_f/\sigma$.  Thus, again, we see that the energy distribution has shifted by the weak value, and has thus served as the pointer for a weak measurement, and again this was a measurement of how much energy was transferred from the photon to the absorber.

To summarize the results, the overall probability for a photon input into the MZ to exit the dark port is $P_\textrm{Dk} = |\alpha|^2/4$, and the energy transferred (nonlocally) when this occurs is $(1-|\alpha|^2)\omega_f$.  The limiting case of $\alpha=1$ is a `classical' obstruction, where the ground state is fully inside the photon's path in arm I, and no energy is exchanged during the interaction-free measurement.  The limiting case $\alpha=0$ is the case that the photon entirely misses the absorber, the dark port never fires, and no energy is transferred.

\section{Discussion}

The new type of weak measurement we are describing has a number of features which distinguish it from the standard case.  It satisfies the general condition for a weak measurement, that after the weak coupling, a single measurement of the pointer energy does not enable one to distinguish any of the photon path or absorber energy degrees of freedom, because the two systems are barely entangled.  Indeed, if this were not the case, then the two distinct energy distributions would fail to interfere at the second beam splitter (because a single measurement of the pointer energy would then constitute a strong projective measurement of the photon path degree of freedom in the MZ), and we could not infer that a dark port detection had produced an interaction-free measurement.

Unlike most weak measurements, the state of the measured system is significantly disturbed by the photon, since the change in energy needed to move it all the way to the orthogonal state still corresponds to a negligible shift of the photon energy distribution.

However, if we follow the two-step process we have developed in the previous sections, we treat this disturbance as occurring first, which creates the pre-selection, $|\psi\rangle$, and increases the energy of the joint system.  Then $|\psi\rangle$ acts on the distribution of the photon, removing the excess energy from the joint system.  In this  second step the two systems become weakly entangled, and we can see by tracing out the energy distribution that $|\psi\rangle$ is barely disturbed by this coupling.  Thus, it is the second step which corresponds to the usual weak measurement scenario, with the special exception that it must also remove the excess energy created by the first step.  Recall that physically, these two steps are mixed together, rather than occurring in this logical sequence, and energy is conserved continuously throughout.

Finally, the weak values obtained by this method are never anomalous, meaning that they are always real, and always fall between the (negative) ground and first excited state energies.  The key to this fact lies in the special way we arrive at the pre-selection $|\psi\rangle$ for energy-based weak measurements.  The pre-selection is always the state of the photon path and absorber, after conditioning on the photon not being absorbed.  This conditioning brings the coefficients $\alpha$ and $\beta$ which are used to define $|g\rangle$ and $|f\rangle$ (and thus $\hat{A}$) into the pre-selection $|\psi\rangle$, whereas in a general weak measurement, $|\psi\rangle$ is independent of the observable $\hat{A}$ being weakly measured.  This special dependence is what results in the well-behaved weak value.  Note that for this pre- and post-selection, the weak values of some other observables are anomalous (nonclassical), which was examined in \cite{elouard2019spooky}.

\textbf{Acknowledgments}:---  This research was supported (in part) by the Fetzer-Franklin Fund of the John E. Fetzer Memorial Trust.  Work by C.E. and A.N.J. was supported by the US Department of Energy (DOE), Office of
Science, Basic Energy Sciences (BES), under Grant No. DE-SC0017890.\\

On behalf of all authors, the corresponding author states that there is no conflict of interest.

\begin{center}
    \textbf{References:}
\end{center}
\bibliographystyle{IEEEtran} 
\bibliography{refs}

\end{document}